\documentclass[11pt,twoside]{article}
\usepackage{asp2006}
\usepackage{epsf}
\usepackage{psfig}
\usepackage{lscape}
\usepackage{natbib}

\begin{document}

\title{The Transverse Proximity Effect in Spectral Hardness}
\author{G\'{a}bor Worseck and Lutz Wisotzki}
\affil{Astrophysikalisches Institut Potsdam, An der Sternwarte 16, D-14482~Potsdam, Germany}

\begin{abstract}
We report on our recent detection of the transverse proximity effect of 4
foreground quasars near Q~0302$-$003 ($z=3.285$) as a local hardness fluctuation
in the spectral shape of the intergalactic UV radiation field, found by
correlating the H~I Ly$\alpha$ absorption with the corresponding He~II
Ly$\alpha$ absorption. We argue that the spectral hardness is a sensitive
physical measure to reveal the influence of quasars onto the UV background over
scales of several Mpc, and that it breaks the density degeneracy hampering the
search for the elusive transverse proximity effect. From the transverse
proximity effect on scales of several Mpc we obtain minimum quasar lifetimes of
$\sim$10--30~Myr.
\end{abstract}

\keywords{quasars: general -- quasars: absorption lines -- intergalactic medium -- diffuse radiation}

\section{Introduction}
After reionization the intergalactic medium (IGM) remains highly ionized due to
the intergalactic UV background generated by quasars and star-forming galaxies
\citep[e.g.][]{haardt96,fardal98}, giving rise to the Ly$\alpha$ forest.

In the vicinity of a UV source the source flux locally
enhances the intensity of the intergalactic UV radiation field. The region
affected by this excess flux will be statistically more ionized than the rest
of the IGM along the line of sight, resulting in a statistically increased
transmission, a radiation-induced 'void' in the Ly$\alpha$ forest.
This so-called proximity effect has been detected towards luminous quasars
\citep[e.g.][]{bajtlik88,scott00}. However, a transverse proximity effect
caused by foreground ionizing sources nearby the line of sight has not been
clearly detected in the H~I Ly$\alpha$ forest
\citep[e.g.][]{liske01,schirber04,croft04}. This could be due to a combination
of systematic effects such as anisotropic radiation of quasars, quasar
variability and large-scale structure.

Intergalactic He~II Ly$\alpha$ $303.78$~\AA~absorption has been detected so far
only towards 6 quasars due to the high probability of intervening optically
thick Lyman limit systems that truncate the flux in the far UV. At $z\ga 3$
mostly very strong He~II absorption is observed
\citep{anderson99,heap00,zheng04b}. At $z\la 3$ absorption becomes patchy
\citep{reimers97,reimers05} with voids ($\tau_\mathrm{HeII}\la 1$) and troughs
($\tau_\mathrm{HeII}>3$) evolving to a He~II Ly$\alpha$ forest at $z\la 2.7$
\citep{kriss01,zheng04,fechner06}.

\begin{table}[!ht]
\caption{\label{qsolist}Quasars near Q~0302-003 with redshifts $z$,
$V$ magnitudes, separation angles $\vartheta$ from Q~0302$-$003 and transverse
proper distances $d_\perp(z)$.}
\smallskip
\begin{center}
{\footnotesize
\begin{tabular}{lcrrrrl}
\tableline
\noalign{\smallskip}
QSO		&Abbr.	&$z$	&$V$	&$\vartheta$	&$d_\perp$	&discovery paper\\
                &	&	&	&[\arcmin]	&[Mpc]		&\\
\noalign{\smallskip}
\tableline
\noalign{\smallskip}
QSO~03022$-$0023&F	&$2.142$&$22.5$	&$3.55$		&$1.77$		&\citet{jakobsen03}\\
QSO~03027$-$0027&E	&$2.290$&$19.9$	&$10.31$	&$5.08$		&\citet{worseck06}\\
QSO~03027$-$0010&B	&$2.808$&$21.7$	&$11.24$	&$5.29$		&\citet{worseck06}\\
Q~0302-D113	&D	&$2.920$&$24.3$	&$4.85$		&$2.26$		&\citet{steidel03}\\
QSO~03020$-$0014&A	&$3.050$&$20.5$	&$6.46$		&$2.97$		&\citet{jakobsen03}\\
Q~0301$-$005	&C	&$3.231$&$17.8$	&$22.89$	&$10.34$	&\citet{barbieri86}\\
\noalign{\smallskip}
\tableline
\end{tabular}
}
\end{center}
\end{table}

Due to the different ionization energies of H~I ($E>13.6$~eV) and He~II
($E>54.4$~eV) the ratio of the intergalactic He~II absorption and the
corresponding H~I absorption indicates the spectral shape of the UV background.
Already low-resolution He~II observations revealed a fluctuating spectral shape
in the voids (hard) and the troughs (soft) \citep[e.g.][]{reimers97}. Recent
observations of the He~II Ly$\alpha$ forest show large spectral fluctuations on
very small scales \citep[e.g.][]{kriss01}. Due to the hard ionizing field
required in the He~II voids that is consistent with the integrated radiation of
a surrounding quasar population, these He~II voids have been interpreted as the
onset of He~II reionization in Str\"{o}mgren spheres around hard He~II
photoionizing sources along or near the line of sight \citep{heap00,smette02}.
\citet{jakobsen03} found a quasar matching to a He~II void at $z=3.05$ towards
Q~0302$-$003, thereby providing the first clear case of the transverse proximity
effect.

Here we summarize our recent paper on the detection of the transverse proximity
effect via the relative spectral hardness of the UV radiation field, described
in \citet{worseck06}. We adopt a flat cosmological model with
$\Omega_\mathrm{m}=0.3$, $\Omega_\Lambda=0.7$ and
$H_0=70$~$\mathrm{km}\,\mathrm{s}^{-1}\,\mathrm{Mpc}^{-1}$.

\section{Observations}
We observed an area of $25\arcmin\times 33\arcmin$ around Q~0302$-$003
($z=3.285$) with the ESO Wide Field Imager at the ESO/MPI~2.2~m Telescope in its
slitless spectroscopic mode \citep{wisotzki01} as part of a survey for faint
quasars near known bright quasars. From the $\sim 800$ slitless spectra recorded
in the range 4200--5800~\AA~we obtained two quasar candidates.

Follow-up spectroscopy confirming these two quasars was performed using FORS2 on
ESO VLT UT1. Table~\ref{qsolist} lists all 6 known quasars at $z>2$ within a
radius $<30\arcmin$ around Q~0302$-$003.

Q~0302$-$003 is one of the few quasars showing intergalactic He~II Ly$\alpha$
absorption and has been observed with HST. We retrieved the STIS spectra of
Q~0302$-$003 obtained by \citet{heap00} at a resolution of $1.8$~\AA~and
re-reduced them.

\section{The Fluctuating Spectral Hardness of the UV Radiation Field}
\begin{figure}[!ht]
\plotone{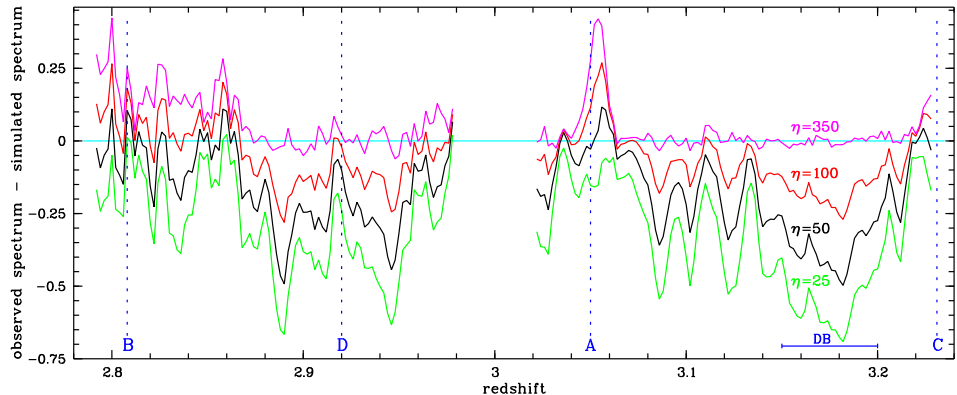}
\caption{\label{diff_eta_asp}Observed vs.~predicted He~II Ly$\alpha$ absorption
in differential re\-presentation for different $\eta$ values. The region
contaminated by geocoronal Ly$\alpha$ emission is not shown. Quasars are marked
as in Table~\ref{qsolist}. Positive (negative) deviations from zero indicate
that $\eta$ has to be smaller (higher) than the assumed $\eta$ of the curve.}
\end{figure}
\begin{figure}[!ht]
\plotone{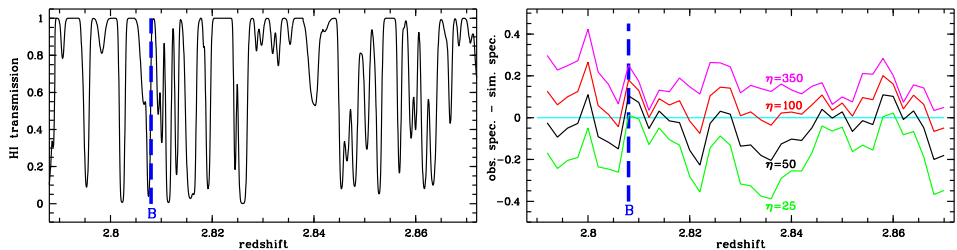}
\caption{\label{diff_eta2_asp}Close-up of the region around QSO~B showing the
H~I Ly$\alpha$ forest as reconstructed from the line list by \citet{hu95}
(left), and the difference between the observed and the simulated He~II data
(right). In the vicinity of QSO~B, $\eta=$25--50 provides the best fit.}
\end{figure}

Observationally the spectral shape of the intergalactic UV radiation field can
be described by the the column density ratio
$\eta=N(\mathrm{HeII})/N(\mathrm{HI})$ of the column densities of He~II and H~I.
Theoretically $\eta\la 100$ for a hard spectrum dominated by quasars, whereas
$\eta\ga 100$ indicates an additional contribution from star-forming galaxies
\citep{haardt96,fardal98}.

At the low resolution of the STIS spectrum of Q~0302$-$003 the He~II absorption
remains unresolved and $\eta$ is not directly accessible. One can still
estimate $\eta$ by comparing the actual He~II spectrum with a simulated
spectrum that is generated using the accurate H~I line parameters from
high-resolution optical spectroscopy and a constant value for $\eta$. We used
the H~I line lists of \citet{hu95} and \citet{kim02}, which jointly cover the
redshift range accessible with STIS in He~II Ly$\alpha$ and simulated the He~II
forest for different values of $\eta$ assuming pure non-thermal broadening of
the lines.

Figure~\ref{diff_eta_asp} shows the resulting He~II spectra degraded to the STIS
resolution as the difference between the observed and the simulated spectrum for
four representative values of $\eta$. The hardness of the UV radiation field
clearly fluctuates as different $\eta$ values match to different redshift
regions. Remarkably, the spectral regions around most of the foreground quasars
are best reproduced by low $\eta$ values which are typical of quasars.

The vicinity of QSO~C is reproduced with $\eta\sim 50$ and from the soft
radiation field in a nearby H~I void \citep[][marked by DB in
Fig.~\ref{diff_eta_asp}]{dobrzycki91} we argue that this void is created by
large-scale structure and not by the quasar. \citet{jakobsen03} found QSO~A
coinciding with a He~II void after \citet{heap00} predicted a quasar at this
redshift due to the inferred hard radiation field. We also find evidence for a
hardness fluctuation near QSO~D. Figure~\ref{diff_eta2_asp} presents an enlarged
view of the redshift region around QSO~B, where we identify an $\eta$ minimum at
the redshift of our foreground quasar. Note that we do not detect the transverse
proximity effect as a void the H~I forest. Instead we find a large-scale
structure filament near QSO~B.
\section{Conclusions}
We have found the transverse proximity effect as a local fluctuation in
spectral hardness near four foreground quasars in the vicinity of Q~0302$-$003.
So far, this effect has mostly been associated with the notion of voids in the
Ly$\alpha$ forest due to the overionized zones around quasars near the line of
sight. We argue that the relative UV hardness is a sensitive \emph{physical}
quantity to search for individual sources of the intergalactic UV radiation
field going beyond the detection of quasars and voids that may be in fact
unrelated.

From the transverse proximity effect detected over Mpc distances we
can obtain lower limits on the lifetimes of the quasars in our sample, yielding
values in the range $\sim$10--30~Myr.

\end{document}